\documentclass{SPAICE}
\usepackage{float}
\def\authorEmail{z.a.rudge@tudelft.nl} % Update with your email
\def\AuthorShort{Z.A. Rudge et al.} % Update with your name

\author[1,2]{Zacharia A. Rudge\thanks{Corresponding author. E-Mail: \authorEmail}}
\author[2]{Giacomo Acciarini}
\author[2]{Dario Izzo}
\author[3]{Dominik Dold}
\author[1]{Moritz Fieback}
\author[1]{Said Hamdioui}
\affil[1]{Delft University of Technology, %Department and 
            Delft, 
            The Netherlands}
            
\affil[2]{European Space Agency, %Department and Organization
            Noordwijk,
            The Netherlands}

\affil[3]{University of Vienna, %Department and Organization
            Wien, 
            Austria}

\title{Improving On-board Shape Models Using Sun Sensors and Implicit Representations}
\begin{document}

% Creates the title and author list automatically for you!
\maketitle

% EDIT HERE
% Please add your abstract here.
\begin{abstract}
% Limit to 200 words. 
% State the problem (limited bandwidth/compute for on-board asteroid shape characterization/limited sensors as well).
% Propose the pipeline (EclipseNets silhouettes to SDF-SIREN).
% Highlight successful reconstruction.
% Highlight potential uses for such a design
We propose an original, end-to-end approach to generate and refine the onboard 3D shape model of irregular small bodies, such as asteroids, using only low-cost Sun sensor eclipse data and coarse a priori knowledge, tailored for resource-constrained missions like CubeSats. Leveraging EclipseNets, a differentiable neural model of eclipse boundaries, our pipeline first refines 2D silhouettes by learning from real-time entry and exit events during spacecraft orbits, avoiding the limitations of traditional imaging or lidar in bandwidth-starved environments. These silhouettes then supervise a compact neural signed-distance field with periodic activations, enabling incremental updates to a high-fidelity implicit geometry representation capable of capturing fine-scale features and concavities. In simulations of the body Eros under realistic orbital dynamics, the method increases silhouette IoU from 0.70 to 0.87 and produces coherent 3D reconstructions, suggesting our approach could find a place in future autonomous architectures for small-body exploration.

\end{abstract}

% EDIT HERE
\section{Introduction}

Small bodies such as asteroids and comets, remnants of the early Solar System, preserve pristine materials and yield critical insights into the chemical and physical processes driving its formation~\cite{SafronovV, yeomans2000small}. Beyond their scientific value, small bodies have also drawn attention for the risk they pose as impact hazards to Earth~\cite{cheng2018aida}.
Consequently, increasing attention has been devoted to their exploration through spacecraft missions, including sample-return campaigns and technology demonstrations~\cite{fujiwara2006rubble, yada2022preliminary, snodgrass201767p, prockter2002near, lauretta2017osiris}. 
As more exploration missions emerge, accurately understanding the dynamical environment around these bodies becomes increasingly important, particularly because of the challenges posed by their irregular shapes, weak gravitational forces, and complex physical characteristics~\cite{scheeres2012orbit}. 

Sizes, materials, and properties of these bodies vary so wildly, few generalizations can be made across different targets. While ground-based techniques like photometry and radar provide valuable preliminary constraints on size, shape, and albedo~\cite{harris1989photometric, millis1989precise}, detailed morphological characterization ultimately requires in-situ observations collected by an orbiting spacecraft~\cite{scheeres1995navigation}.

The irregular geometry of small bodies strongly affects their local dynamical environment. Shape determines the gravitational field, constrains internal mass distribution, and influences orbital stability and spacecraft maneuvering. It also affects illumination and eclipse conditions, which modulate solar radiation pressure, a dominant perturbation in the weak-gravity regime of small bodies. Accurate shape estimation is therefore fundamental for mission design, guidance, navigation, control, and operational safety, as well as for optimizing scientific observation strategies.

Currently, high-fidelity shape models are generated by combining multiple images or lidar scans, processed using photogrammetry, stereophotoclinometry~\cite{gaskell2023stereophotoclinometry, terik2022shape}, or shape-from-silhouette techniques~\cite{bandyonadhyay2019silhouette}. While producing highly detailed 3D reconstructions, these methods are computationally intensive and depend on continuous ground-based processing. This workflow limits real-time adaptability, which is increasingly problematic given the rise of small, low-cost satellites with constrained onboard resources and communication bandwidth~\cite{starek2015spacecraft, nesnas2021autonomous}.

To address these challenges, autonomous onboard shape reconstruction has become an active area of research~\cite{dennison2023autonomous, villa2022autonomous}. Machine learning approaches, in particular, offer the potential to efficiently generate and refine 3D models from sparse or partial observations~\cite{chen2024asteroid, tang2025asteroid, bandyonadhyay2019silhouette}. Among these, occupancy networks and other implicit surface representations are especially promising for small-body missions, as they can represent complex geometries and be updated incrementally with new measurements.
However, existing learning-based methods typically require RGB imagery or depth data and assume sufficient downlink capacity for ground processing. No prior work has demonstrated end-to-end onboard shape reconstruction using only eclipse events from Sun sensors, a capability that could reveal useful for low-cost missions.

In this work, we leverage EclipseNets~\cite{acciarini2025eclipsenets, biscani2022reliable}, pre-trained on rough a priori knowledge of the asteroid’s shape, to bootstrap the reconstruction process. As the spacecraft orbits the target and collects additional silhouette information (derived, for instance, from points of entrance and exit in eclipse regions detected with a simple Sun sensor), this data is used to iteratively train an occupancy network. The resulting workflow produces a continuously improving 3D representation of the asteroid, enabling real-time, onboard shape reconstruction. This approach is particularly suited to the challenges of modern small-body exploration, supporting autonomous navigation, hazard assessment, and efficient scientific operations on missions with limited computational and communication resources.

The proposed pipeline operates in three stages. First, a weak EclipseNet model is trained offline from coarse a priori asteroid knowledge to approximate eclipse boundaries. During the mission, Sun sensor eclipse entry and exit events are collected and used to refine this model in situ. The refined EclipseNet generates silhouettes for arbitrary viewpoints, which are then converted into orthographic rays and used to train a neural signed-distance field representation of the asteroid. This final representation produces a continuously improving 3D model suitable for onboard navigation and scientific analysis.
%These techniques can generate complex three-dimensional shapes from saprse observations, and progressively refine the 

% - General fluff sentence on the topic of on-board AI
% - Motivation for autonomous, on-board asteroid characterization.
% - Benefits of federated learning on smallsats with limited budget (in energy and cost) for sensors and compute.
% - 
% - Core narrative: Deploying a generalized "weak" model and refining it in-situ with event-based data.

\section{Methodology}
% This section details our full pipeline from eclipsenets 
% short intro to the idea of converting binary silhouettes to 3d shapes

\subsection{Silhouette Generation from a Priori Knowledge}

The first stage of our approach generates 2D silhouettes of the asteroid based on a priori knowledge of the asteroid shape. We leverage EclipseNets, a neural implicit model trained to predict the occurrence of solar eclipses as a function of spacecraft position and Sun direction. Given the spacecraft position $\pmb{r}$ and Sun unit vector $\hat{\pmb{s}}$, EclipseNet, a neural network with learnable parameters $\pmb{\theta}$, outputs a scalar eclipse function $e_{\pmb{\theta}}(\pmb{r}, \hat{\pmb{s}})$: a signed distance function (SDF) defined as negative inside the shadow, positive outside, and zero along the silhouette boundary.
EclipseNet employs SIREN periodic activation functions~\cite{sitzmann2020implicit}, enabling it to capture high-frequency details of complex asteroid contours. The model is initially trained using coarse prior knowledge of the asteroid’s shape before the spacecraft visit, following the approach detailed in previous work~\cite{acciarini2025eclipsenets}. The EclipseNet used in this work comprises three hidden layers of width 32 with sinusoidal activation functions and contains 2,369 trainable parameters, requiring approximately 9.5 kB of FP32 parameter memory.

\subsection{In-Situ Model Refinement}
% does this go before the other stuff?
% - Describe the baseline "weak" model training 
% - Explain the in-situ refinement process using the limited (1 million samples) event sample dataset.
% - Generation of data
% 4000 orbits, 500 days with one satellite
% - Note the orbital mechanics constraints that limit the views.

The second stage focuses on iterative onboard refinement of the asteroid’s 3D shape using eclipse events collected during the mission. Instead of relying on high-resolution imaging or lidar, we exploit precise entry and exit points of the eclipse, derived from spacecraft position and Sun direction. These events are collected via a Sun sensor instead of a traditional camera and serve as ground truth data for continuous training and refinement. 

To generate realistic data from a Sun sensor, we simulate multiple orbital trajectories with randomized initial conditions. Orbital radii are varied between 1.2 and 4 times the semi-major axis of the Eros ellipsoid ($\approx$ 32km), while inclinations are uniformly sampled from 0° to 90°. Our experiments simulate 25 virtual satellites over 100-day orbits, yielding /~1.4 million eclipse events, roughly comparable to the 2.1 million samples per epoch used in offline EclipseNet training and thus enabling a good comparison of refinement efficacy. Although these orbits are not optimized and no control is applied, the resulting eclipse events serve to demonstrate the potential of our method under random conditions.

The spacecraft dynamics are modeled in a body-centered frame as~\cite{biscani2022reliable}:

\begin{align}
\begin{split}
    \dot{\pmb{r}} &= \pmb{v}, \\
    \dot{\pmb{v}} &= -G \sum_{j=0}^{N} \frac{m_j}{|\pmb{r} - \pmb{r}_j|^3} (\pmb{r} - \pmb{r}_j) - 2 \boldsymbol{\omega} \times \pmb{v} \\
    &\quad - \boldsymbol{\omega} \times (\boldsymbol{\omega} \times \pmb{r}) - \eta \nu(\pmb{r})\hat{\mathbf{s}}(t),
\end{split}
\end{align}

where $\pmb{r}$ and $\pmb{v}$ are the spacecraft position and velocity, $m_j$ and $\pmb{r}_j$ represent mascon masses and positions modeling the asteroid’s irregular gravity field, $\boldsymbol{\omega}$ is the asteroid’s angular velocity, and $\eta$ quantifies acceleration due to solar radiation pressure ($10^{-3}$ m/s$^2$ outside eclipse). The eclipse factor $\nu(\pmb{r})$ is zero in shadow and one otherwise. Since the eclipse factor is smooth and differentiable via EclipseNet, a Taylor integrator can accurately track eclipse transitions~\cite{biscani2022reliable}.

\subsection{Neural Implicit Representation of 3D Shapes}
We approach the 3D reconstruction problem using a neural network with a SIREN-SDF at its core, bypassing the limitations of traditional view-based algorithms. Methods such as visual hulls~\cite{Laurentini1994TheVH}, Gaussian splatting, and NeRFs~\cite{kerbl3Dgaussians, mildenhall2020nerf} are unsuitable for purely binary silhouettes, as they often fail to capture concavities, overestimate volume, or require RGB and depth information. Our implicit neural representation of the investigated body is given by the following expression:

%Alternative view-based algorithms like visual hulls \cite{Laurentini1994TheVH} or gaussian splatting and NeRFs \cite{kerbl3Dgaussians, mildenhall2020nerf} either fail to capture concavities, produce significantly larger shapes than necessary or require RGB and depth information to operate, making them unsuitable for binary silhouettes. %Algorithms such as visual hull  will contain the original reconstructed shape and have consistent silhouettes, but can fail to capture concavities and generally produces shapes significantly larger than the true object shape. Other methods include Gaussian splatting and NeRFs , both of which are generally designed for views of scenes that include RGB and depth information. 
%All of these do not map well to our targeted objective of an accurate and improved 3D representation of a planetary body informed by purely binary silhouettes.  
% Explain the SDF-SIREN approach. 

\begin{equation}\label{eq.sdf-siren}
\phi_{\pmb{\theta}'}(\pmb{x}) = \underbrace{\|\pmb{x}\|-r_0}_{\text{sphere SDF init}} + \underbrace{f_{\pmb{\theta}'}(\pmb{x})}_{\text{SIREN residual}}
\end{equation}

\cref{eq.sdf-siren} evaluates whether a given point is inside or outside the body, with positive values of $\phi_{\pmb{\theta}'}(\pmb{x})$ corresponding to being outside the body, and negative ones to being inside. The sphere prior provides an initial topology, while the residual is trainable (with parameters $\pmb{\theta}'$) to deform this macro-shape. The SIREN activation enables the neural network to model high frequency details.  %without positional encoding??
To convert this expression to occupancy, we use the following sigmoid expression:

\begin{equation}\label{eq.sigmoid}
o_{\pmb{\theta}'}(\pmb{x})=\sigma(-\alpha\phi_{\pmb{\theta}'}(\pmb{x}))
\end{equation}

The sigmoid function $\sigma$ converts a given output of the SIREN-SDF to a value bounded by the range $[0, 1]$, where $\alpha$ controls the sharpness of the transition. After applying this expression, an output of 0 (1) denotes high confidence of this point being outside (inside) the body. 

During training, we generate a number of views (represented by binary silhouettes of the body) using the earlier stage's EclipseNet. Training is performed for 1,000 epochs. At each optimization step, eight silhouettes (or views) are sampled, with 384 rays selected per silhouette and 64 points evaluated along each ray. This corresponds to 3,072 rays and 196,608 ray-sampled SDF evaluations per optimization step. The SIREN-SDF used in this work comprises six hidden layers of width 192 and contains 186,241 trainable parameters, requiring approximately 745 kB of FP32 parameter memory.

For each view, a random set of points on this 2D binary silhouette is converted into a set of (parallel) rays. Such a ray ($r(t) = o + td$) is clipped by training bounding box to produce $(t_{near}, t_{far})$. The binary silhouette can be considered as an orthographic camera, with each pixel in the silhouette being equivalent to a parallel camera ray casting towards the asteroid. 

\begin{figure}[h]
    \centering
    \includegraphics[width=1\linewidth]{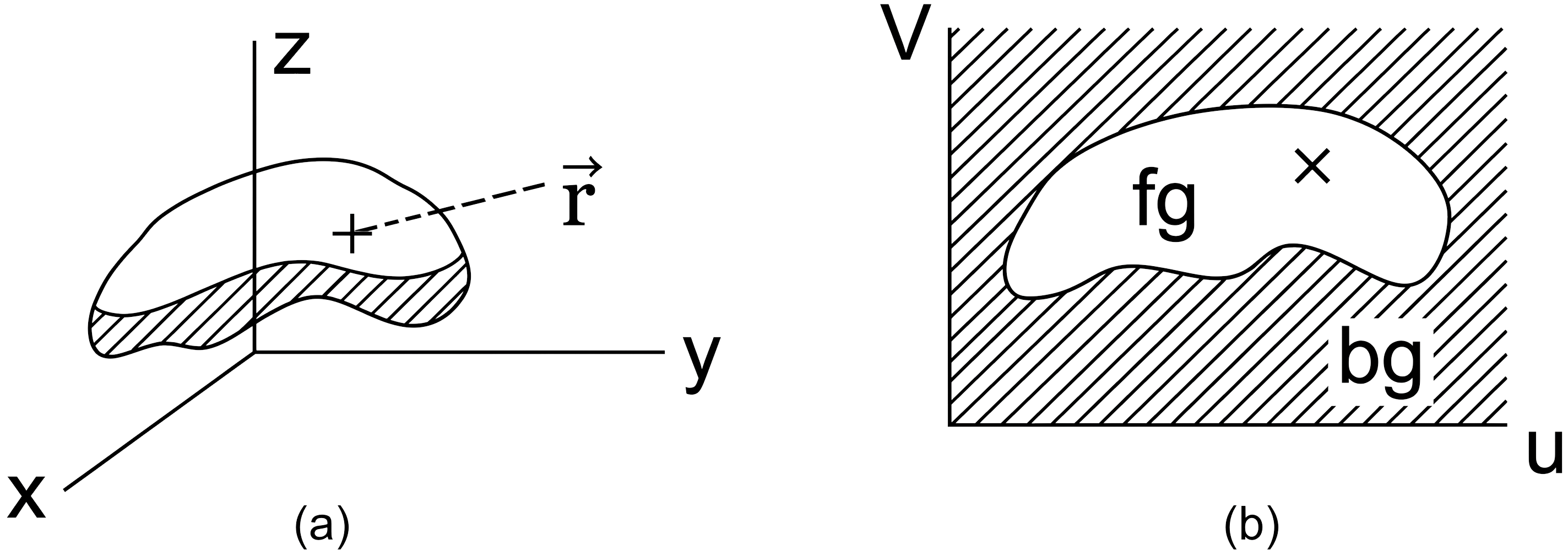}
    \caption{Schematic representation of the point to orthographic ray conversion, (a) shows the target body as generated by the SIREN-SDF, (b) shows a silhouette generated by the refined eclipseNet.}
    \label{fig:ray_conversion}
\end{figure}

This method of sample conversion is illustrated in \cref{fig:ray_conversion}. In this figure, (a) gives a schematic representation of the target body as generated by the SIREN-SDF. $+$ denotes where the ray $r(t)$ intersects the body. (b) shows a sampled silhouette as generated by the refined eclipseNet model. $fg$ and $bg$ denote the foreground (silhouette) and background (empty space) of the silhouette, with $X$ showing the sampled point from which the ray is converted.
%new figure here?

As the 2D silhouettes from which we are training do not contain depth information, we must collapse individual point $o_{\pmb{\theta}'}(\pmb{x})$ evaluations down to a single 2D prediction (on the plane of the view) by calculating \cref{eq.phit}, 

\begin{equation}\label{eq.phit}
P_{hit}=\max_i o_i
\end{equation}

The final loss is then computed by comparing the $P_{hit}$, produced by evaluating multiple points along the chosen ray, and the ray's associated 0 (outside) or 1 (inside). %this sentence is a bastard

To summarize, EclipseNets is used to produce 2D views, we obtain rays from points in these views. Then these rays are evaluated using the SIREN-SDF, the loss is then evaluated and we train to reduce the mismatch between this evaluated value and the original 2D view's value on that point. %It also encodes graded supervision around boundaries by converting the binary silhouettes to smooth gradient slopes informing the network how far it missed by, and in which direction it needs to direct the surface. 
Lastly, the training also includes scheduled curvature regularization to encourage smooth global geometry and stable normals. This is scheduled to start halfway through the training, as regularization that is too eager will prevent the model from learning the macro-shape and will limit its extremities too much. 
During inference the threshold of $o_{\pmb{\theta}'}(\pmb{x})$ at which we consider a space occupied is swept for using the resulting best "Intersection over Union" (IoU) between $[0, 1]$.

% Clarify the continuous domain vs. binary classification

% Detail the projection mechanism (u, v coordinates -> 3D infinite rays which are trimmed to a certain range). 
% Explain the loss terms: Visual Hull loss, Coverage loss, Curvature regularizer (and its scheduling).

% Full Pipeline figure (eventually)
% \begin{figure*}[t]
%     \centering
%     \includegraphics[width=.95\textwidth]{figures/pipeline_architecture.pdf}
% 	\caption{Full pipeline from Sun Sensor Events to Geodesy extraction, including the EclipseNets to SDF-SIREN conversion.}
% 	\label{fig:pipeline_architecture} 
% \end{figure*}

\section{Results}
%The results were evaluated in two distinct ways: Firstly, the ability for a weakly trained EclipseNet to be refined using in-situ obtained eclipse event samples. Secondly, the performance of the SIREN-SDF-based model to accurately reconstruct the targeted 3D shape using the information produced by the refined EclipseNet.
% Every paper requires a results section.
%simple intro on what results will be presented

\subsection{Contour Reconstruction Accuracy}
% Quantitative metrics comparing the reconstructed Eros (and others) to ground truth.
% Show state in weak and in refined states 
First, we evaluate the 2D contour accuracy using the following three metrics: Intersection over Union (IoU) quantifies the spatial overlap between the predicted silhouette and the ground truth, where a value of 1.0 indicates perfect alignment. The XOR error measures the area of disagreement (sum of false positives/negatives). Finally, the Fill metric denotes the proportion of predicted foreground pixels. The results are show in \cref{tab:metric_summary}, where we can see that from weak to refined there is a strong jump in both IoU and fill. 

\begin{table}[ht]
\begin{center}
\resizebox{\columnwidth}{!}{%
\begin{tabular}{lcccc} 
\toprule
\textbf{Model} & \textbf{Events Seen} & \textbf{Fill} & \textbf{IoU} & \textbf{XOR} \\
\midrule
Ground Truth & N/A & 0.062981 & 1.000000 & 0.000000 \\
Refined & 1376256 & 0.062332 & 0.867514 & 0.007698 \\
Weak & 0 & 0.043861 & 0.695617 & 0.019119 \\
\bottomrule
\end{tabular}%
}
\caption{Metric comparison summary between the weak baseline, the refined model, and ground truth.}
\label{tab:metric_summary}
\end{center}
\end{table}

\begin{figure*}[!h]
    \centering
    \includegraphics[width=.95\textwidth]{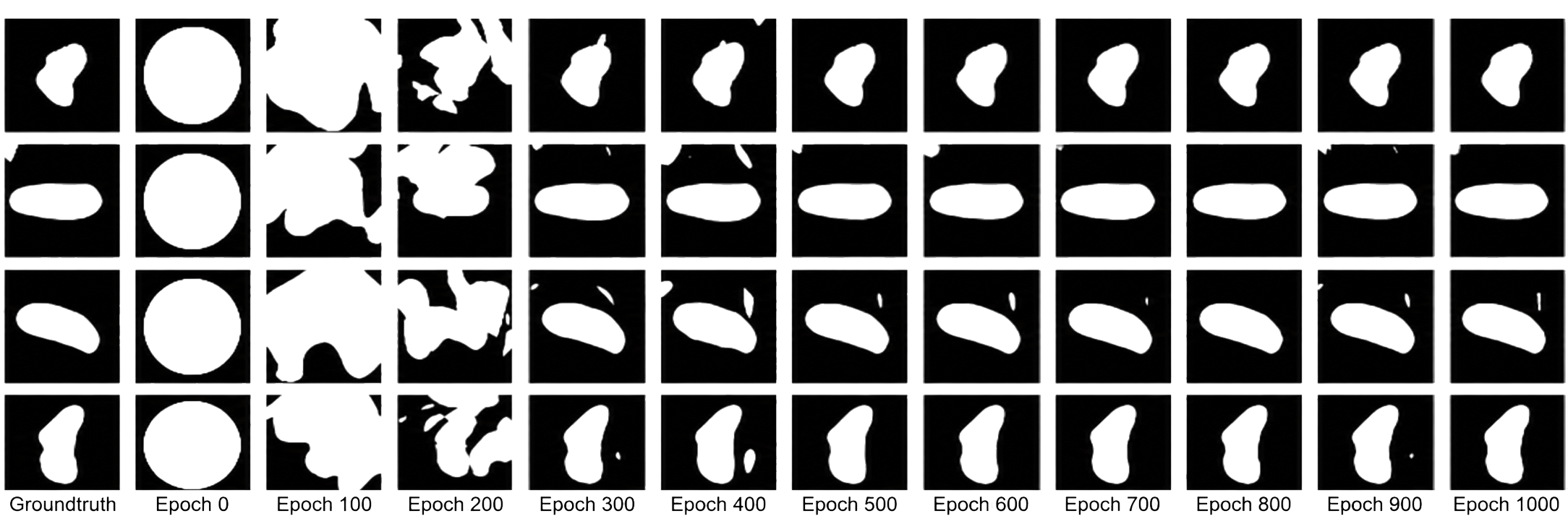}
    \caption{The process of training the occupancy network based on the provided EclipseNet silhouettes. Curvature regularization starts halfway through the training process. Groundtruth shapes are shown in the first column.}
    \label{fig:trained_occupancy}
\end{figure*}

\begin{figure}[!h]
    \centering
    \includegraphics[width=.99\columnwidth]{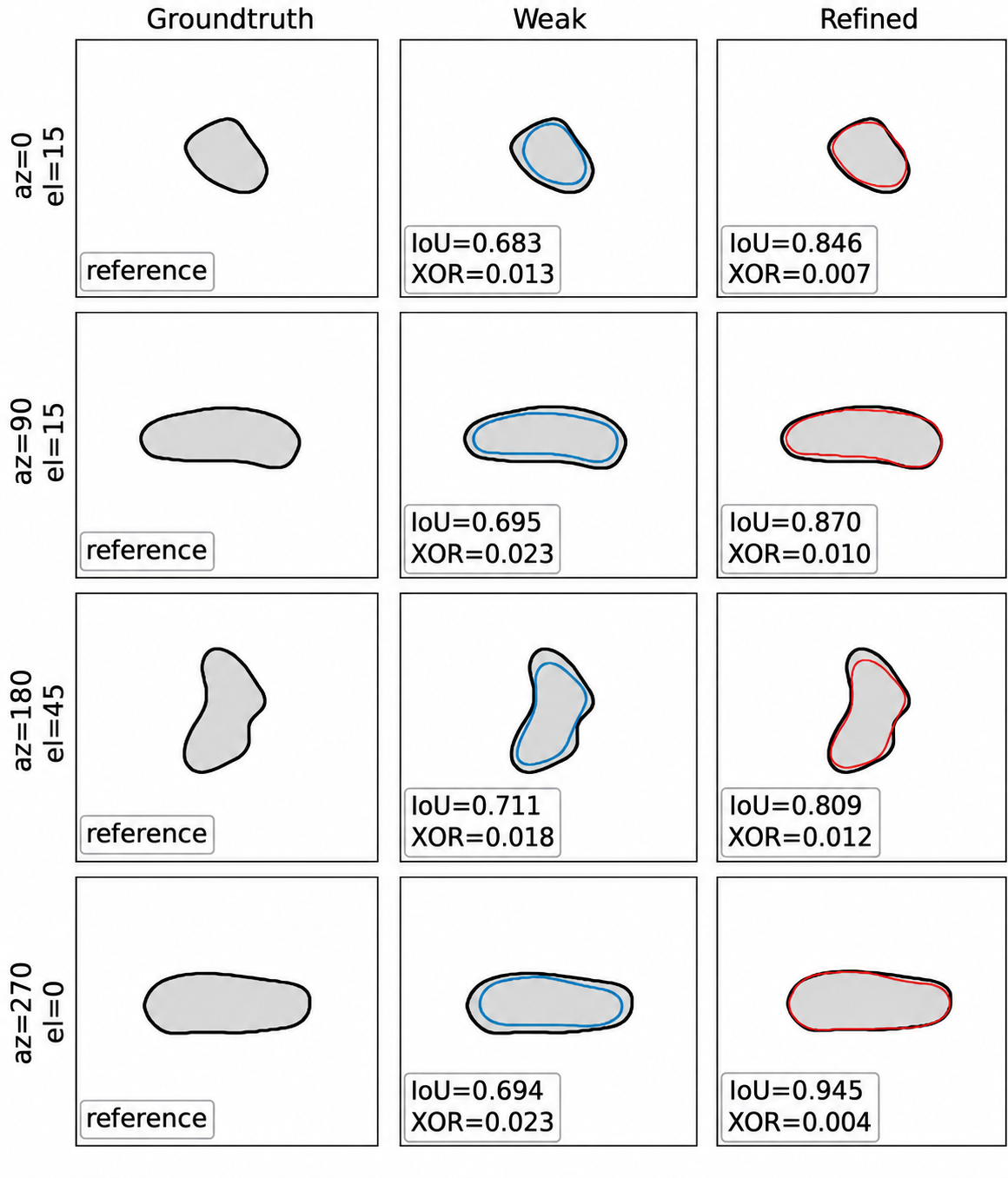}
    \caption{Plot of the ground truth (first column), the initial representation of the target body upon Earth departure (second column) and the refined model (third column).}
    \label{fig:weak_refined}
\end{figure}

This matches what can be visually observed in \cref{fig:weak_refined}.
The first column provides the ground truth silhouettes (also shown in the other columns as the black outline). The second column displays the predictions from the weakly trained EclipseNet upon initial deployment, where structural underfilling is visible. The third column demonstrates the output of the refined model after the network has processed the additional in-orbit samples. The additional in-orbit samples alleviate the initial overfilling, closely aligning the model with the ground truth. We can also see that though the model's fitting is significantly improved, on the view of $azimuth=180^\circ$, $elevation=45^\circ$, the boundary doesn't precisely match and is the least improved of all views shown. This is possibly be related to the number of samples (the in-situ samples are slightly over 1 million, and one training epoch of the offline training is 2.1 million samples), or to the distribution of samples (the angles contained in the eclipse events). 

\subsection{3D Model Reconstruction Accuracy}

In \cref{fig:trained_occupancy}, the visual evolution of the volume represented by the occupancy network across the training epochs is shown (ending at training epoch 1000). The network initially diverges from the sphere prior before rapidly converging to a plausible target shape, with subsequent curvature regularization removing anomalous occupancies.

\begin{figure}[!h]
    \centering
    \includegraphics[width=.95\columnwidth]{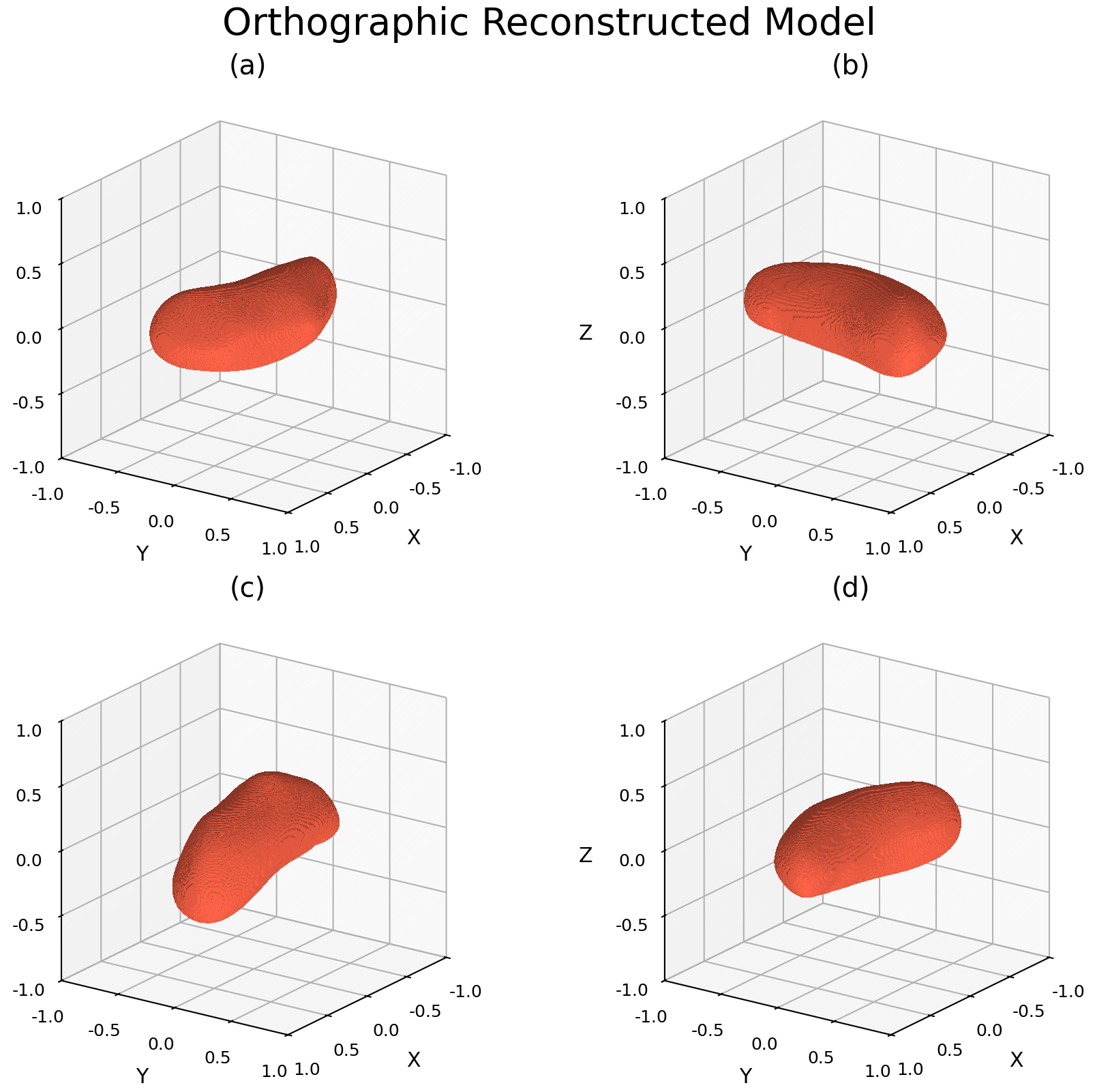}
    \caption{3D render of the final reconstructed shape, based on the trained occupancy network. Four angles (a, b, c and d) are presented.}
    \label{fig:final_model}
\end{figure}

The complete reconstructed shape is shown as a polyognal 3D model in \cref{fig:final_model} from several angles. This model can then be used for navigation, asteroid landing, or geodesy on-board. The code and data to reproduce the results in this paper is available on GitHub \cite{github_repo}.

\section{Discussion}

This work presents a pipeline for in-situ reconstruction of asteroid morphology using only Sun sensors and neural implicit representations. The results demonstrate that meaningful geometric information can be extracted from sparse event-based measurements and used to iteratively refine an onboard model of the target body. Starting from a weak prior representation encoded in an EclipseNet, the incorporation of in-orbit eclipse entry and exit events significantly improves silhouette reconstruction accuracy. These refined silhouettes can subsequently be used to train an implicit signed distance representation, producing a plausible three-dimensional model of the asteroid.

The quantitative results indicate that the refinement stage substantially improves silhouette agreement with the ground truth. Visually, the refined silhouettes capture the morphology of the target body and reduce the systematic underfilling present in the weak baseline model.

Although the reconstruction problem is inherently ill-posed when relying solely on binary silhouettes, the use of a sphere prior combined with a SIREN residual model to capture higher-frequency geometric features seem to be able to infer a smooth, coherent and globally consistent surface that satisfies the observed silhouettes constraints. The resulting shape reconstruction preserves silhouette consistency across viewpoints while producing a physically plausible volume. 

From an operational perspective, the proposed pipeline is particularly relevant for resource-constrained spacecraft missions. As this pipeline bypasses the specialized imaging hardware and extensive ground processing required by traditional techniques, which may be incompatible with small spacecraft platforms. In contrast, the method presented here relies entirely on Sun sensor measurements, which are among the most common and lightweight sensors available on spacecraft. 

%By exploiting eclipse entry and exit events, the spacecraft can acquire geometric information about the asteroid without requiring a dedicated imaging hardware.

Despite these promising results, several limitations must be acknowledged. First, reconstruction quality depends heavily on the distribution of observed eclipse events, dictated by spacecraft trajectory and illumination geometry. %Limited viewing angles fail to sufficiently constrain certain surface regions, leading to lower accuracy from less represented viewpoints. 
Future work should investigate ad-hoc orbit design strategies to maximize silhouette diversity. Second, the pipeline assumes the coarse prior shape initializing EclipseNet approximates the true geometry. Incorporating uncertainty-aware priors or adaptive regularization strategies may improve robustness against poor initial estimates. Sun-sensor noise's impact on the in-situ training has also not yet been taken into account. 
Finally, further qualitative assessment of the reconstructed 3D Shape of its geometric accuracy in various metrics also requires further development.

Overall, the results demonstrate that eclipse-based observations combined with neural implicit representations provide a viable pathway for autonomous small-body morphology reconstruction. By leveraging simple sensors and onboard learning, spacecraft could incrementally refine their understanding of the target body's geometry during operations. This capability has potential applications in autonomous navigation for future asteroid exploration missions.

% Every paper requires a discussion section.
% - Address limitations: the necessity for balanced, shuffled datasets during in-situ refinement.
% Address limitations in terms of orbit
% Address that 

\begin{acknowledgments}
Zacharia A. Rudge acknowledges funding from the European Space Agency's Open Space Innovation Platform (contract number 4000140774). Dominik Dold was supported by the Horizon Europe's Marie Skłodowska-Curie Actions (MSCA) Project 101103062 (BASE).
\end{acknowledgments}

\printbibliography
\addcontentsline{toc}{section}{References}

\end{document}